# Laterally-confined two-dimensional electron gases in self-patterned LaAlO$_3$/SrTiO$_3$ interfaces

M. Foerster,[1] R. Bachelet,[1] V. Laukhin,[1,2] J. Fontcuberta,[1] G. Herranz,[1,*] and F. Sánchez[1,*]

[1] Institut de Ciència de Materials de Barcelona (ICMAB-CSIC), Campus de la UAB, 08193 Bellaterra, Catalonia, Spain

[2] Institució Catalana de Recerca i Estudis Avançats (ICREA), 08010 Barcelona, Catalonia, Spain

A bottom-up process has been used to engineer the LaAlO$_3$/SrTiO$_3$ interface atomic composition and locally confine the two-dimensional electron gas to lateral sizes in the order of 100 nm. This is achieved by using SrTiO$_3$(001) substrate surfaces with self-patterned chemical termination, which is replicated by the LaAlO$_3$ layer, resulting in a modulated LaO/TiO$_2$ and AlO$_2$/SrO interface composition. We demonstrate the confinement of the conducting interface forming either long-range ordered nanometric stripes or isolated regions. Our results demonstrate that engineering the interface chemical termination is a suitable strategy towards nanoscale lateral confinement of two-dimensional high-mobility systems.

A few years ago, high-mobility conduction was discovered at the interface between LaAlO₃ (LAO) and SrTiO₃ (STO), two robust insulators with wide energy bandgaps –$E_g$ = 5.6 eV and 3.2 eV, respectively–.[1] This conductive layer is confined to a few unit cells around the interface[2,3,4] forming a two-dimensional electron gas (2DEG)[5] that is observed when LAO films with thickness above 3 – 4 unit cells are grown on STO crystals, whereas for thinner LAO layers the interfaces remain insulating.[6] On the metallic side, electron mobility above $10^3$ cm²/Vs has been reported.[5] Interestingly, it has been found that the electronic state of the LAO/STO interface can be modified with photons[7] or it may be toggled between metallic and insulating states either by writing charges on the LAO films with atomic force microscopy (AFM) tips[8,9,10,1112] or by adsorption of surface charges using polar solvents.[13] Whereas high-mobility conduction is observed for LAO layers grown on TiO₂-terminated STO, the SrO-terminated interfaces are highly resistive.[1,14] Experimentally, it is observed that a gradual increase of the fractional SrO coverage of the used STO substrate from pure TiO₂- to full SrO-terminations, realized by deposition of a SrO submonolayer on TiO₂-terminated STO substrates,[1,15] gives way to a steady increase of the resistivity and a transition from metallic to insulating behavior.[15] This observation, illustrating the extreme sensitivity of the interface properties against their chemical atomic structure, could provide a strategy alternative to top-down processes[16,17,18,19] towards the patterning of electronic properties

While the formation of the 2DEG at the LAO/STO interface is now well established and it has inspired novel electronic devices, lateral nanoscale confinement of these systems could open new possibilities. In the experiments described above, the surface termination is gradually modified from TiO₂ to SrO terminations by random partial coverage of a sub-monolayer of SrO prior to the LAO deposition and thus it does not allow to produce ordered terminations at the interface. Here, we exploit our ability to obtain large-scale STO(001) surfaces with well-oriented self-assembled patterns of SrO- and TiO₂- terminations by high temperature treatments,[20,21,22] and using these nanostructured substrates as templates for deposition of LAO. The film grows replicating the chemical termination pattern of STO, yielding a conspicuous pattern of LaO/TiO₂ and AlO₂/SrO atomic stacking sequences at the interface. We will show that the electronic properties of the interfaces are laterally modulated. Long ribbons of conducting 2DEGs, some few hundred nanometer wide, or embedded 2DEGs nanoregions in insulating matrix can be obtained ad-hoc, thus illustrating the potential of surface-termination self-ordering in perovskite oxides.

Prior to film deposition, STO substrates were treated in a dedicated furnace by choosing annealing temperature and time to obtain distinct patterns of the chemical terminations. Detailed information about the self-formation of chemical terminations patterns is reported elsewhere. [20, 21, 22] Here, three STO(001) substrates were treated at 1000 ºC (for 2 hours) and 1300 ºC (for 2 and 6 hours) to yield different surface chemical termination patterns. Subsequently, LAO thin films were grown by pulsed laser deposition ($\lambda$ = 248 nm) monitored by high pressure reflection high energy electron diffraction (RHEED). Before deposition, the substrates were warmed up from room temperature to the deposition temperature (850 ºC) in an oxygen partial pressure $P_{O2}$ = 0.1 mbar. Then, LAO was grown under $P_{O2} = 10^{-4}$ mbar at 1 Hz repetition rate, with laser pulse energy of 22 - 26 mJ. The LAO thickness, 5 or 6 ML, was controlled in-situ by monitoring the RHEED specular spot intensity (with incidence of electrons along STO[100] at a glancing angle of 1.5-3º). At the end of the deposition, samples were cooled down in oxygen rich atmosphere to minimize the formation of oxygen vacancies: [2,6, 23] $P_{O2}$ = 0.3 mbar from T = 850 ºC to 750 ºC and $P_{O2}$ = 200 mbar from T = 750 ºC to room temperature, including a dwelling time of 1 hour at 600 ºC. The surface morphology was analyzed by scanning electron microscopy (SEM) and AFM working in dynamic mode (images were analyzed using the WSxM software[24]), and the spatial distribution of the chemical terminations was scrutinized by phase-lag imaging. The interface resistivity was measured in van der Pauw geometry by contacting the LAO top via ultrasonic bonding with Al wire. The contact geometry was set in order to inject the current along two perpendicular directions, one along the direction of the terrace steps and the atomic composition patterns –as determined from AFM imaging– and the other normal to them. The distance between contacts was of the order of a few millimeters. From these experiments, we obtained the $R_{//}$ and $R_{\perp}$ resistances, and the resistance anisotropy A was determined following the analytical method proposed by O. Bierwagen et al.[27] The temperature dependence of the sheet carrier density and electronic mobility were obtained in Hall experiments in van der Pauw configuration.

First we discuss interfaces prepared on STO substrates that were treated by a relatively soft annealing (1000 ºC, 2 hours). The overall morphology of these treated substrates is flat with meandering steps and terraces mostly $TiO_2$-terminated and coexisting with small SrO-terminated regions not resolved with standard AFM [see Refs. 20, 21, 22]. The RHEED pattern of the annealed substrate just before LAO deposition (Figure 1a) shows diffraction spots on the $0^{th}$ Laue circle, confirming the overall flatness. The intensity oscillations of the specular spot during deposition of LAO (Figure 1b) indicate a layer-by-layer growth mode, which permitted stopping the growth after completing exactly six monolayers (ML). We note that 6 unit cells are above the critical thickness for 2DEG formation. The RHEED pattern at the end of the growth (Figure 1c) is similar to that of the used substrate. The AFM topographic image in Figure 1d shows a flat film with terraces around 350 nm wide separated by meandered steps. The pattern of substrate chemical-terminations is expected to be replicated in the LAO film by the layer-by-layer growth mechanism. However, the minority LaO terminated small regions are not resolved in the topographic (Figure 1d) or phase-lag (not shown here) AFM images. Interestingly, high resolution SEM detecting either secondary or backscattered electrons permits distinguishing them clearly. Although the physical origin causing the contrast is not evident, the SEM images map accurately the pattern (see the backscattered electrons image in Figure 1e). The minority termination is in small regions, typically a few tens of nanometers in size. Its inter-spacing depends on the distance to the substrate steps, in agreement with the surface diffusion mechanism proposed in ref. [22]. The expected interface, mostly $LaO/TiO_2$ with small and sparse regions of insulating $AlO_2/SrO$ clusters, is sketched in Figure 1f. Thus, overall, the interface should be conductive and its properties essentially those of the conventional metallic LAO/STO interface. Figure 1g displays the temperature dependence of $R_{//}$ and $R_{\perp}$, where $R_{//}$ and $R_{\perp}$ are the resistance measured along the substrate steps and perpendicular to them, respectively; the data in this Figure show that this interface is isotropic. Note that at room temperature the sheet resistances are $R_{\perp} \approx R_{//} \approx 32$ k$\Omega$/sq., whereas at the lowest temperature, $R_{\perp} \approx R_{//} \approx 0.4$ k$\Omega$/sq., which are comparable to the reported values in literature. [5, 6] We note that previous reports seemed to point to a strong influence of steps on the in-plane angular dependence of transport of LAO/STO interfaces grown on $(LaAlO_3)_{0.3}(Sr_2AlTaO_3)_{0.7}$ substrates.[25] In the present case, with well visible steps between wide terraces (Figure 1d), the transport is isotropic, thus suggesting that the interface topology (steps) does not induce any transport anisotropy.

A second STO substrate was treated at 1300 ºC for 2 hours. In this case, the main effect of the thermal treatment is the coalescence of the existing SrO termination fraction next to the terraces steps, yielding a regular pattern of two-terminated chemical composition evident in the topographic (Figure 2a) and phase-lag (Figure 2b) AFM images. Terraces are around 300 nm wide, with the pattern of distinct termination oriented along the steps, $TiO_2$-termination being majority with SrO-terminated stripes located next to the steps.[20,21, 22] The intensity of the RHEED specular spot (see in Figure 2c the pattern of the substrate before deposition) was monitored to stop the deposition after growth of six ML (Figure 2d). The RHEED pattern of the deposited film (Figure 2e) displays well-defined spots on the $0^{th}$ Laue circle, indicating atomically-flat and single-crystalline surface. Backscattered electrons SEM and phase AFM images (Figures 2f and 2g, respectively), as well as topographic AFM images (see supplementary material[Error! Marcador no definido.]), confirm that the LAO surface reproduces the underlying substrate surface ordering. Thus, $AlO_2$ is the majority termination, with minority LaO termination placed next to straight steps in elongated regions less than 100 nm wide and up to around 3 μm in length. The replication of the chemical termination pattern from STO substrate to LAO film implies that the atomic stacking at the interface presents the same pattern, and thus the conducting $LaO/TiO_2$ interface is separated by elongated regions with insulating $AlO_2/SrO$ interface, as sketched in Figure 3a. Naturally, such pattern induces anisotropy in the transport properties, as observed in the electrical transport measurements (Figure 3b). In order to determine the real resistance anisotropy and obtain the intrinsic sheet resistances we considered the effect of current path redistribution on the transport anisotropy using the model reported in Refs. [[26],[27]]. At room temperature, the sheet resistance was $R_\perp \approx 310$ kΩ/sq. and $R_{//} \approx 60$ kΩ/sq., whereas at low temperature we recorded $R_\perp \approx 2.8$ kΩ/sq. and $R_{//} \approx 0.55$ kΩ/sq., respectively (Figure 3b). Thus, the sheet resistance anisotropy ratio varied within the range $A \approx 5 - 7$ for all the temperature range (Figure 3b, lower panel). Note that the sheet resistance along the steps ( $R_{//}$ ) has a value very similar to the one measured for the isotropic interface discussed above. In addition, from Hall measurements we extracted the mobility and sheet carrier density (Figure 3c). We have assumed that the transport anisotropy comes essentially from the directional dependence of the mobility, whereas the sheet carrier density is independent on current injection path (parallel or perpendicular to steps) (see Ref. [25]). We obtain then a sheet carrier density, which is weakly dependent on temperature, $\sim 1.1 \times 10^{13}$ cm$^{-2}$, whereas the electronic mobility measured for current injected along the patterns is significantly larger than across them ($\mu_{//} \approx 1200$ cm$^2$/Vs *versus* $\mu_{\perp} \approx 200$ cm$^2$/Vs at low temperature). The comparison of these results with those described above obtained on substrates without ordering of the minority chemical termination, demonstrates that the high resistance anisotropy is originated from the large-scale patterning of the interface atomic composition.

Finally, a third substrate was prepared by annealing at T = 1300 ºC for 6 hours. As in the previous case, a regular surface morphology of terrace steps appears with two chemical terminations, but here with SrO being the majority phase. The long treatment causes SrO enrichment of the surface by Sr diffusion from the bulk of the single crystal towards its surface, and results in a characteristic terrace and step morphology (Figures 4a and 4b), as described in reference [20]. The width of the SrO terraces and the elongated TiO$_2$ regions is around 500 nm and 200 nm, respectively. The RHEED pattern in Figure 4c, recorded before deposition, shows spots on the Laue circle similar to the other samples, but also additional spots at half positions compatible with 2x2 reconstruction. A recent study[28] has reported emergence of 2x2 reconstruction in SrO-terminated STO after heating under the same pressure (0.1 mbar oxygen) as used here to heat up the substrate. A LAO film with thickness of five ML was grown on this surface (see the RHEED intensity oscillations in Figure 4d), with the pattern at the end of the deposition (Figure 4e) showing only integral order reflections. Backscattered electron SEM images (Figure 4f) show the chemical termination pattern of the substrate is again replicated in the film, thus with minority AlO$_2$ elongated regions within the majority LaO-terminated surface. The chemical termination pattern at the film surface is also evidenced in AFM phase (Figure 4g) and AFM topographic (see supplementary material[**Error! Marcador no definido.**]) images. The replication of the stacking sequence during the growth implies that this sample presents elongated regions, unconnected LaO/TiO$_2$ across the film/substrate interface (see the sketch in Figure 4h). These regions, containing highly conducting interfaces, are surrounded by the majority insulating $AlO_2/SrO$ thus rendering the sample macroscopically insulating. Indeed, the transport measurements indicate a resistance beyond our limiting value (~100 MΩ). Comparison of the sketches in Fig. 4h and 3a helps to visualize the differences among these two samples.

To summarize, we have shown that using nanostructured $SrTiO_3$ substrates as templates we can laterally confine conductive regions of nanometric size in the two-dimensional electron gas at the $LaAlO_3/SrTiO_3$ interface. This goal has been demonstrated with two pattern geometries: i) highly ordered conducting ribbons of $LaO/TiO_2$ terminations separated by narrow stripes of insulating $AlO_2/SrO$ interfaces, giving rise to a pronounced in-plane anisotropy of carrier mobility and resistivity; and ii) oriented elongated nanometric regions with non-percolating conducting $LaO/TiO_2$ interface embedded in a matrix with insulating interface. This bottom-up fabrication process provides a strategy for large-scale control of the transport properties of $LaAlO_3/SrTiO_3$ interfaces at the nanoscale. Finally, in the light of the extreme sensitivity of the interface properties to the chemical composition, one could speculate about the role of small inhomegeneities of the interface atomic stacking sequence –not resolved by atomic force microscopy– that could eventually trigger the coexistence of magnetic order and superconductivity recently observed at the $LaAlO_3/SrTiO_3$ interface. [29,30]

Financial support by the Spanish Government (Projects MAT2008-06761-C03, MAT2011-29269-C03, and CSD2007-00041) and Generalitat de Catalunya (2009 SGR 00376) is acknowledged.

Figure 1. (Color online) (a) RHEED pattern taken along the [100] direction of the STO substrate treated at 1000 °C for 2 hours. (b) Intensity of the specular spot during deposition of 6 ML of LAO (note that the intensity of the incident electrons was manually increased at time from 157 to 181 s), and pattern at the end of the deposition (c). (d) AFM topographic and (e) backscattered electrons SEM images of the film. (f) Sketch of the atomic stacking sequences at the interface. (g) Temperature dependence of the resistance with current injected along and across the surface steps, showing isotropic transport.

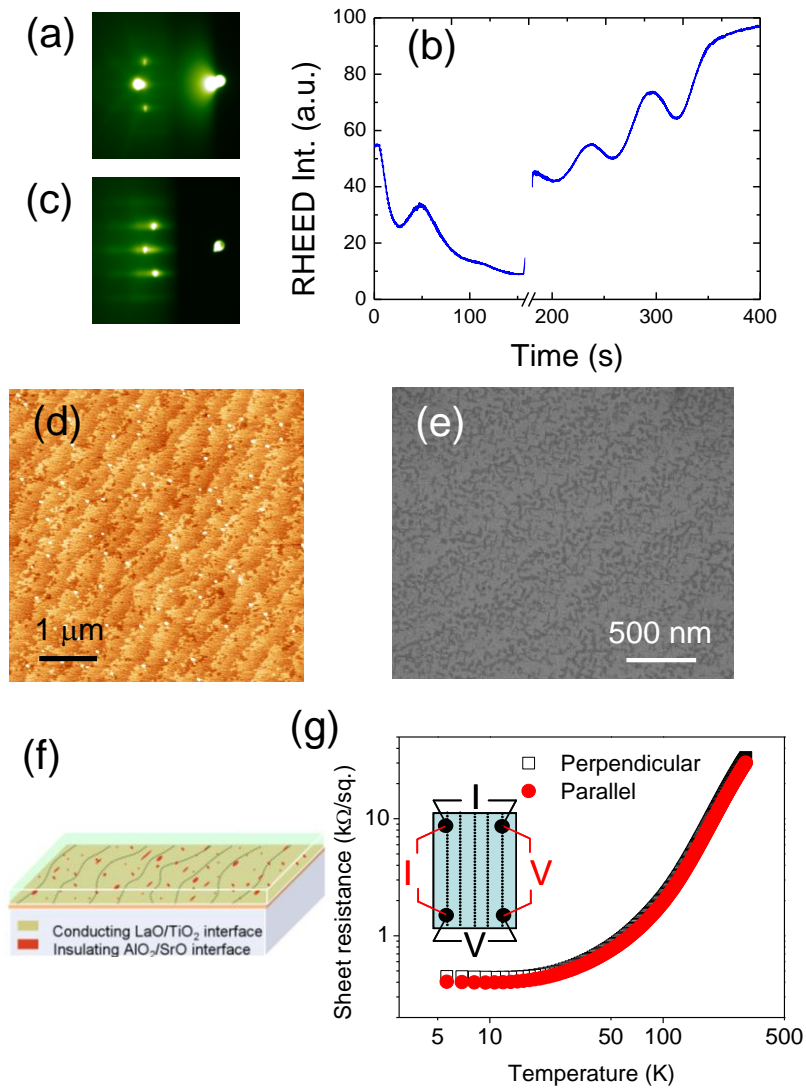

Figure 2. (Color online) AFM topographic (a) and phase (b) images, and RHEED pattern taken along the [100] direction (c) of the STO substrate treated 2h at 1300 ºC (the insets are closer views). (d) Intensity of the specular spot during deposition of 6 ML of LAO, with a zoom of the last oscillations in the inset, and pattern of the film at the end of the deposition (e). (f) Backscattered electrons SEM and (g) AFM phase images of the film.

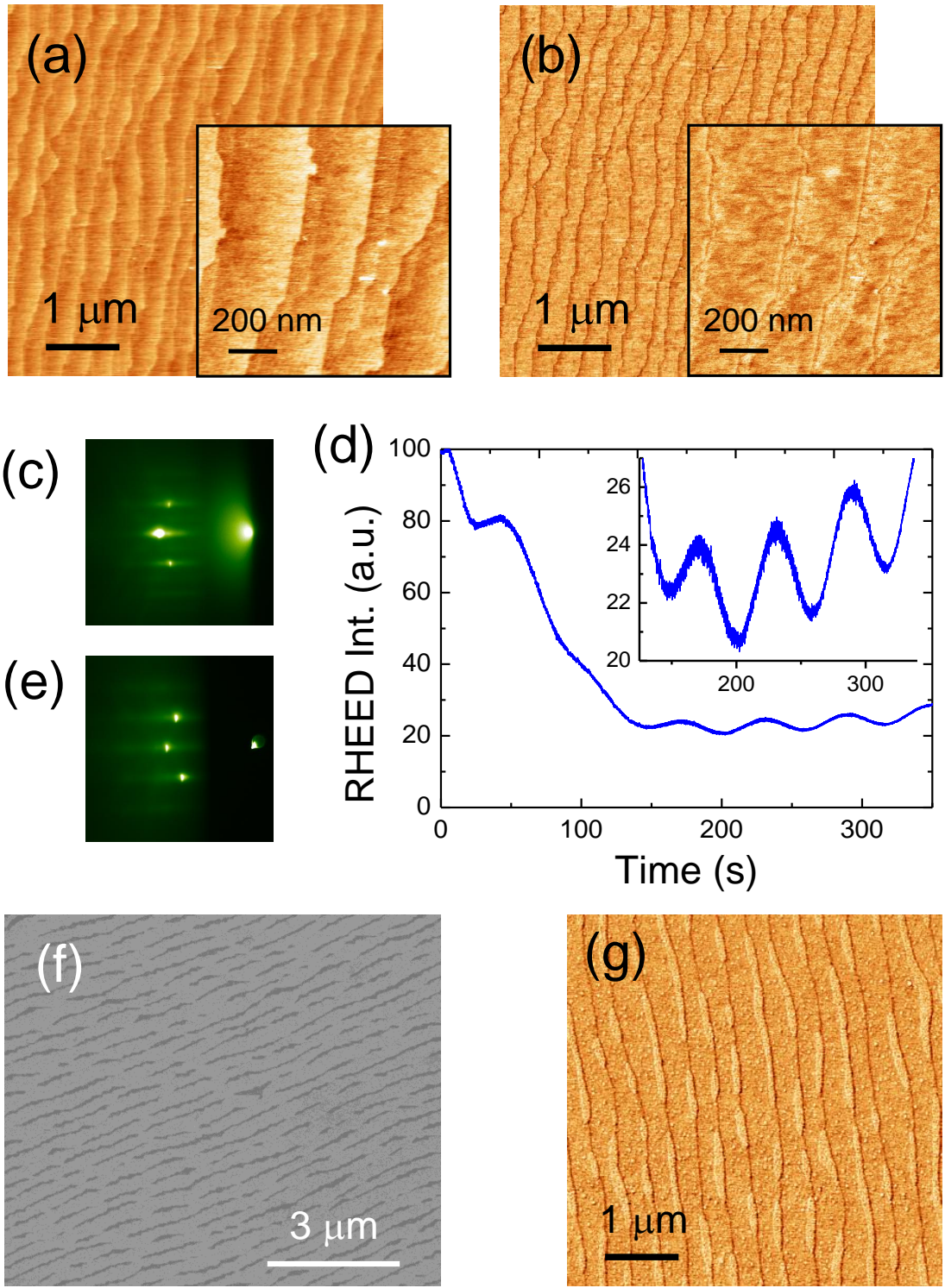

Figure 3. (Color online) (a) Sketch of the atomic stacking sequences at the interface of the LAO film on STO treated at 1300 ºC for 2 hours. (b) Temperature dependence of the sheet resistance with current injected along ($R_{//}$) and across ($R_{\perp}$) the surface steps (top) and resistance anisotropy (bottom). (c) Temperature dependence of the mobilities μ// and $\mu_{\perp}$. The inset shows the sheet carrier density.

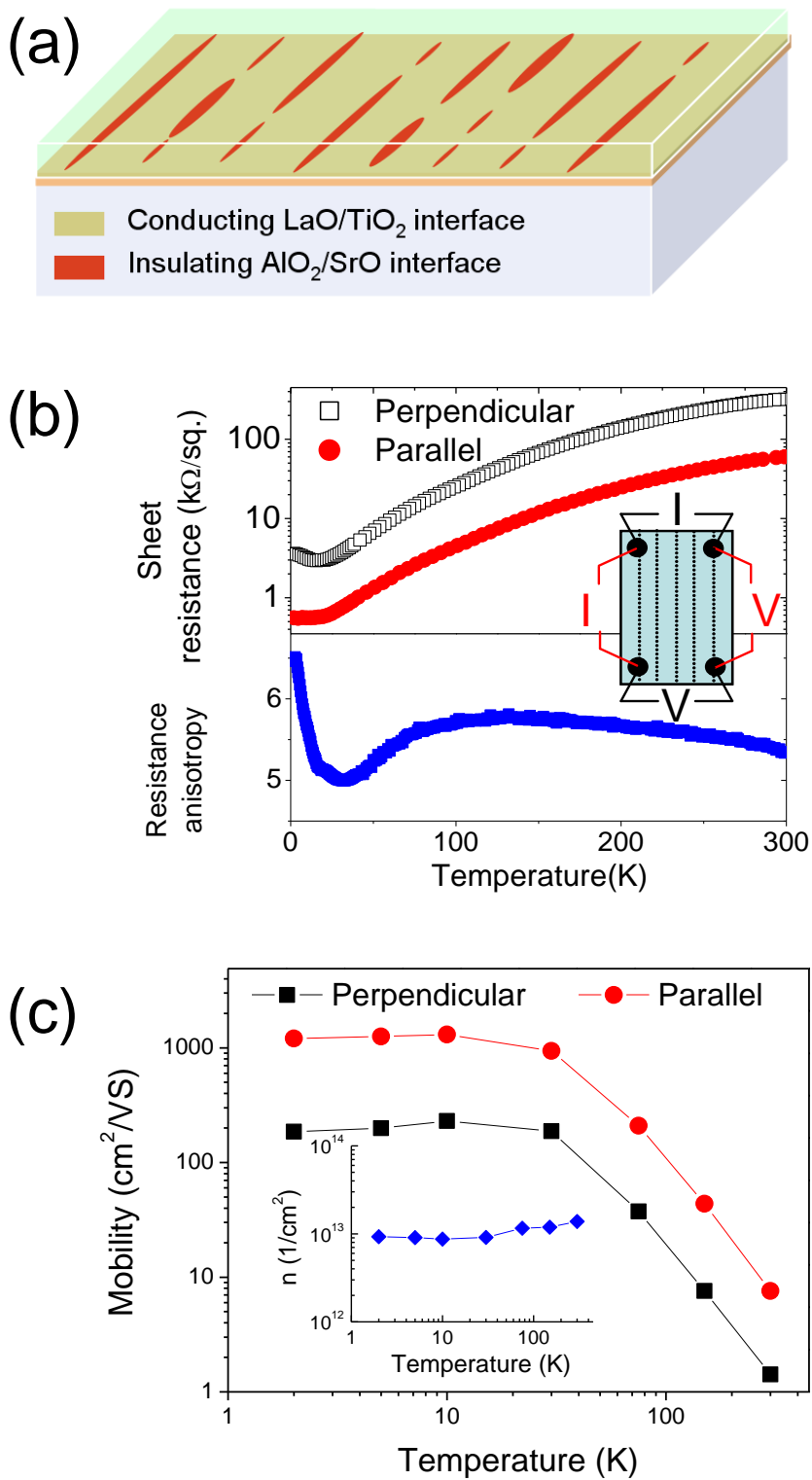

Figure 4. (Color online) AFM topographic (a) and phase (b) images, and RHEED pattern (c) taken along the [100] direction of the STO substrate treated 6 h at 1300 ºC. (d) Intensity of the specular spot during growth of 5 ML of LAO, and pattern of the film at the end of the deposition (e). (f) Backscattered electrons SEM and (g) AFM phase images of the film. (h) Sketch of the atomic stacking sequences at the LAO/STO interface.

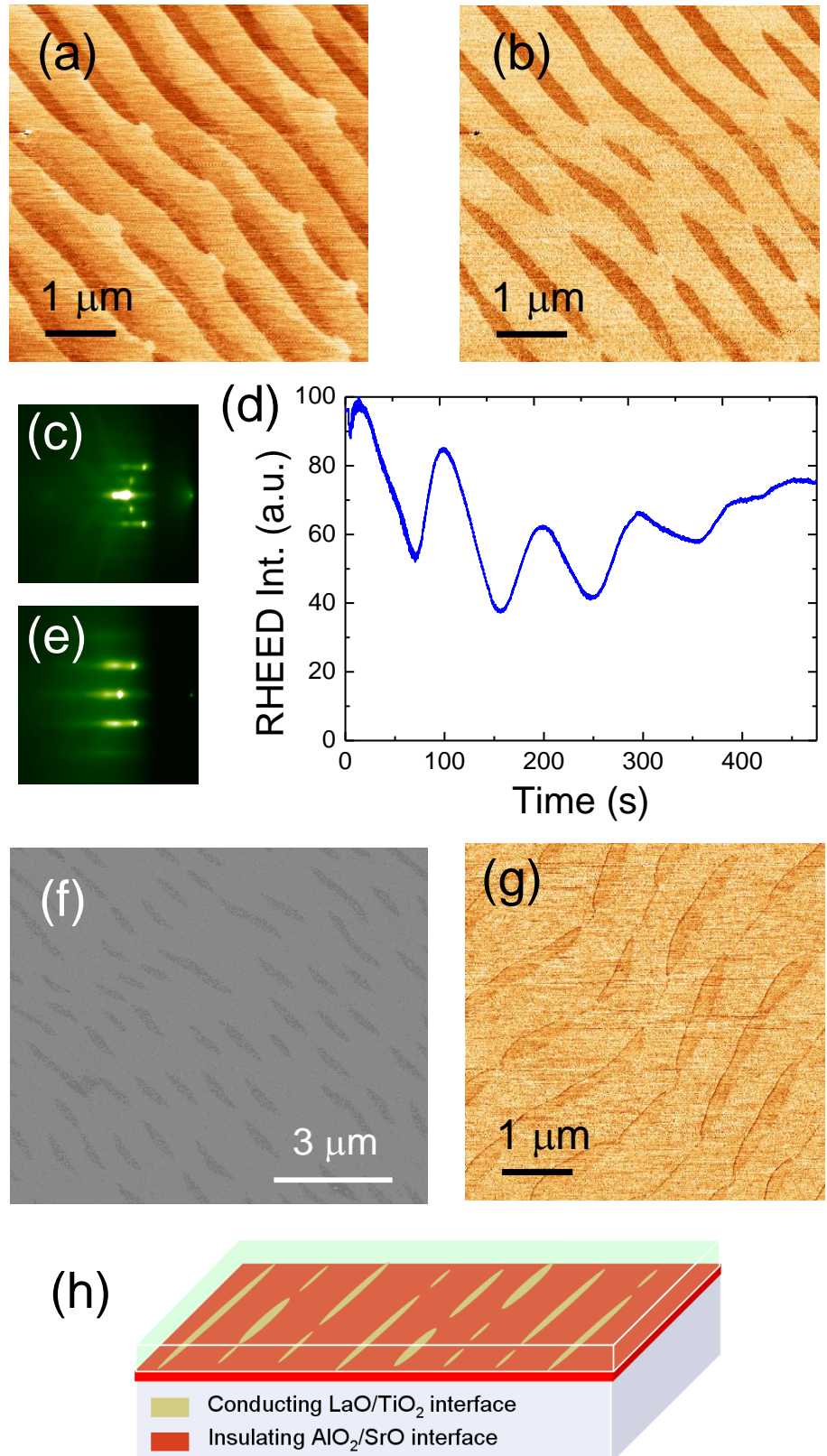

**Supplementary material**

**Figure S1:** AFM topographic images of the LAO films on STO treated at 1300 ºC during 2h (a) and 6h (b).

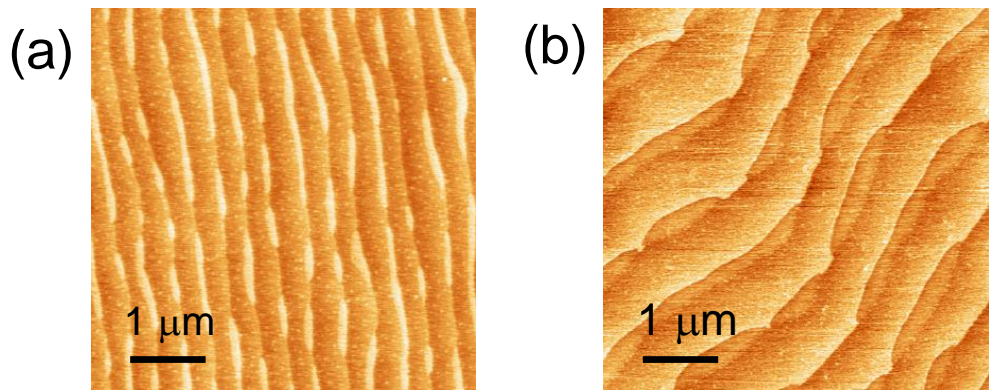

[1] A. Ohtomo and H. Y. Hwang, Nature **427**, 423 (2004).

[2] M. Basletic, J. L. Maurice, C. Carrétéro, G. Herranz, O. Copie, M. Bibes, E. Jacquet, K. Bouzehouane, S. Fusil, and A. Barthélémy, Nat. Mater. **7**, 621 (2008).

[3] O. Copie, V. Garcia, C. Bödefeld, C. Carrétéro, M. Bibes, G. Herranz, E. Jacquet, J. L. Maurice, B. Vinter, S. Fusil, K. Bouzehouane, H. Jaffrès, and A. Barthélémy, Phys. Rev. Lett. **102**, 216804 (2009).

[4] N. Reyren, S. Thiel, A. D. Caviglia, L. F. Kourkoutis, G. Hammerl, C. Richter, C. W. Schneider, T. Kopp, A. S. Rüetschi, D. Jaccard, M. Gabay, D. A. Muller, J. M. Triscone, and J. Mannhart, Science **317** 1196 (2007).

[5] A. D. Caviglia, S. Gariglio, C. Cancellieri, B. Sacépé, A. Fête, N. Reyren, M. Gabay, A. F. Morpurgo, and J. M. Triscone, Phys. Rev. Lett. **105**, 236802 (2010).

[6] S. Thiel, G. Hammerl, A. Schmehl, C. W. Schneider, and J. Mannhart, Science **313**, 1945 (2006).

[7] A. Tebano, E. Fabbri, D. Pergolesi, G. Balestrino, and E. Traversa, ACS Nano **(**2012), article published ASAP, 10.1021/nn203991q

[8] C. Cen, S. Thiel, G. Hammerl, C. W. Schneider, K. E. Andersen, C. S. Hellberg, J. Mannhart, and J. Levy, Nat. Mater. **7**, 298 (2008)

[9] C. Cen, S. Thiel, J. Mannhart, and J. Levy, Science **323**, 1026 (2009)

[10] Y. W. Xie, C. Bell, T. Yajima, Y. Hikita, and H. Y. Hwang, Nano Lett. **10**, 2588 (2010)

[11] Y. W. Xie, C. Bell, Y. Hikita, and H. Y. Hwang, Adv. Mater. **23**, 1744 (2011)

[12] P. Irvin, Y. Ma, D. F. Bogorin, C. Cen, C. W. Bark, C. M. Folkman, C. B. Eom, and J. Levy, Nat. Photonics **4**, 849 (2010)

[13] Y. W. Xie, Y. Hikita, C. Bell, and H. Y. Hwang, Nat. Comms. **2**, 494 (2011)

[14] M. Huijben, A. Brinkman, G. Koster, G. Rijnders, H. Hilgenkamp, and D. H. A. Blank, Adv. Mater. **21**, 1665 (2009)

[15] J. Nishimura, A. Ohtomo, A. Ohkubo, Y. Murakami, and M. Kawasaki, Jpn. J. Appl. Phys. **43**, L1032 (2004)

[16] B. Förg, C. Richter, and J. Mannhart, Appl. Phys. Lett. **100**, 053506 (2012)

[17] N. Banerjee, M. Huijben, G. Koster, and G. Rijnders, Appl. Phys. Lett. **100**, 041601 (2012)

[18] C. W. Schneider, S. Thiel, G. Hammerl, C. Richter, and J. Mannhart, Appl. Phys. Lett. **89**, 122101 (2006)

[19] C. Bell, S. Harashima, Y Kozuka, M. Kim, B. G. Kim, Y. Hikita, and H. Y. Hwang, Phys. Rev. Lett. **103**, 226802 (2009)

[20] R. Bachelet, F. Sánchez, F. J. Palomares, C. Ocal, and J. Fontcuberta, Appl. Phys. Lett. **95**, 141915 (2009).

[21] R. Bachelet, F. Sánchez, J. Santiso, C. Munuera, C. Ocal, and J. Fontcuberta, Chem. Mater. **21**, 2494 (2009)

[22] M. Paradinas, L. Garzón, F. Sánchez, R. Bachelet, D.B. Amabilino, J. Fontcuberta, and C. Ocal, Phys. Chem. Chem. Phys. **12**, 4452 (2010)

[23] G. Herranz, M. Basletić, O. Copie, M. Bibes, A. N. Khodan, C. Carrétéro, E. Tafra, E. Jacquet, K. Bouzehouane, A. Hamzić, and A. Barthélémy, Appl. Phys. Lett. **94**, 012113 (2009)

[24] L. Horcas, R. Fernandez, J.M. Gomez-Rodriguez, J. Colchero, J. Gomez-Herrero, and A. M. Baro, Rev. Sci. Instrum. **78**, 013705 (2007)

[25] P. Brinks, W. Siemons, J. E. Kleibeuker, G. Koster, G. Rijnders, and M. Huijben, Appl. Phys. Lett. **98**, 242904 (2011)

[26] S. H. Simon, Phys. Rev. Lett. **83**, 4223 (1999)

[27] O. Bierwagen, R. Pomraenke, S. Eilers, and W. T. Masselink, Phys. Rev. B **70**, 165307 (2004)

[28] A. Fragneto, G. M. De Luca, R. Di Capua, U. Scotti di Uccio, M. Salluzzo, X. Torrelles, T. L. Lee, and J. Zegenhagen, Appl. Phys. Lett. **91**, 101910 (2007)

[29] L Li, C. Richter, J. Mannhart, and R. C. Ashoori, Nat. Physics **7**, 762 (2011)

[30] J. A. Bert, B. Kalisky, C. Bell, M. Kim, Y. Hikita, H. Y. Hwang, and K. A. Moler, Nat. Physics **7**, 767 (2011)